\documentclass[prb,superscriptaddress,preprint,showpacs,showkeys]{revtex4}

\usepackage{graphics}
\usepackage{graphicx}

\begin{document}

\bibliographystyle{apsrev}

\title{Tight-binding study of the influence of the strain on the electronic properties
of InAs/GaAs quantum dots}

\author{R. Santoprete}
\thanks{Corresponding author. Tel.: +55-21-25627336; Fax: +55-21-25627368}
\email[Email address: ]{rsantop@if.ufrj.br.}
\affiliation{Instituto de F\'{\i}sica, Universidade Federal do Rio de
Janeiro, Cx. Postal 68528, 21941-972 Rio de Janeiro, Brazil}
\affiliation{Fritz-Haber-Institut der Max-Planck-Gesellschaft, Faradayweg 4-6, D-14195 Berlin-Dahlem, Germany}
\author{Belita Koiller}
\author{R. B. Capaz}
\affiliation{Instituto de F\'{\i}sica, Universidade Federal do Rio de
Janeiro, Cx. Postal 68528, 21941-972 Rio de Janeiro, Brazil}
\author{P. Kratzer}
\affiliation{Fritz-Haber-Institut der Max-Planck-Gesellschaft, Faradayweg 4-6, D-14195 Berlin-Dahlem, Germany}
\author{Q. K. K. Liu}
\affiliation{Bereich Theoretische Physik, Hahn-Meitner-Institut, Glienicker Str. 100, D-14109 Berlin, Germany}
\author{M. Scheffler}
\affiliation{Fritz-Haber-Institut der Max-Planck-Gesellschaft, Faradayweg 4-6, D-14195 Berlin-Dahlem, Germany}

\date{\today}

\begin{abstract}
We present an atomistic investigation of the influence of strain on the electronic properties of 
quantum dots (QD's) within the empirical $s p^{3} s^{*}$ tight-binding (ETB) model with interactions up to 2nd nearest 
neighbors and spin-orbit coupling. Results for 
the model system of capped pyramid-shaped InAs QD's in GaAs, with supercells containing
$\sim 10^{5}$ atoms are presented and compared with previous empirical pseudopotential results.
The good agreement shows that ETB is a reliable alternative for an atomistic treatment.
The strain is incorporated through the atomistic valence force field model. The ETB treatment allows for the effects of bond length and bond angle deviations from the ideal
InAs and GaAs zincblende structure to be selectively removed from the electronic-structure calculation, giving quantitative information 
on the importance of strain effects on the bound state energies and on the physical origin of the spatial elongation of the wave functions.
Effects of dot-dot coupling have also been examined to determine the relative weight of both strain field
and wave function overlap.
\end{abstract}



\maketitle

\pagebreak

\section{Introduction}
\label{introduction}

Nanometer-size semiconductor quantum dots (QD's) have attracted scientific interest due 
to their potential applications in optoelectronic devices as well as because of their peculiar properties such as
the self-assembly, tunability, narrow size distribution and large Coulomb blockade effects. \cite{bimberg1}
The size and shape of the Stranski-Krastanow growth of InAs QD's on GaAs (001) reported by different authors 
vary, depending on the epitaxial method and on the growth conditions. Different sizes of QD's, pyramidal or dome 
shapes with side facets oriented along different directions, \cite{ruvimov,nabetani,moison,solomon,lee-lowe}
truncated cone \cite{shumway} and pyramids \cite{fry} with 
nonuniform Ga incorporation in the nominally InAs QD's have been reported.
The driving force for the formation of 
such structures is the relief of the elastic energy associated with a dislocation-free, epitaxial structure (the InAs/GaAs lattice mismatch is 7\ \%). The strain 
distribution is not uniform, so accurate electronic models should include the effects of such nonuniformity.

Theoretical models currently employed in the study of the electronic properties of QD's can be generally divided 
into macroscopic or microscopic.
Examples of macroscopic models are the one-band effective mass approximation \cite{marzin,grundmann,vasanelli} 
and the multi-band {\bf k} $\cdot$ {\bf p} models. \cite{pryor,stier} Microscopic models 
are based on the empirical pseudopotential method \cite{zunger1} and on the empirical tight-binding (ETB) method. 
\cite{saito2,saito3,saito,allan1,lee,lee2,bryant,klimeck,dicarlo}
Microscopic models provide an atomistic treatment, as required for a more realistic description of smaller
heterostructures.
Here, the effects of inhomogeneous strain follow directly by taking into account deviations
 of the atomic positions from the ideal InAs and GaAs bulk structures. 
The empirical pseudopotential treatment potentially offers the most accurate description of the electronic properties
of QD's. 
On the other hand, the ETB method may offer a faster alternative, and it is more transparent with respect to the analysis of results in term of the modified chemical bonding present in and around quantum dots.

Despite their potential strength, not many ETB calculations are available regarding capped strained InAs QD's.
The only published results are obtained by modeling the dot by a spherical cluster with dangling bonds saturated 
by hydrogen \cite{allan1,lee,lee2,bryant} (this approximation is valid in the limit of nanocrystals embedded in a 
material with a very wide gap) or by a pyramidal dot with uncovered surfaces. \cite{saito}

In this work we explore the ETB method for evaluating and analyzing the electronic structure of InAs QD's. Our aim here is to:

\begin{enumerate}
\item Study the reliability of the ETB scheme for the treatment of semiconductor nanostructures. We consider a square-based pyramidal InAs QD
with \{101\} side facets embedded in GaAs. Since this geometry has been previously adopted by several authors, 
\cite{pryor,stier,zunger1,saito} the reliability of our results is assessed through comparison with previous 
studies.
\item Investigate how the strain affects the electronic properties of QD's. So far, such investigations
have been limited to spherical \cite{williamson} or elliptical \cite{fu} dots. In the first case, the influence of the strain was estimated by comparing free standing with GaAs embedded quantum dots. However the surface dangling bonds in the free-standing dot were passivated by a fictitious material with a band gap much larger than the GaAs gap, giving raise to a much larger confining effect for electrons and holes inside the dot. Therefore in the comparison not only the different strain configurations played a role, but also the different band offsets.
In the second case a comparison of results of differently strained dots was made, without however including any bond angle deformation.
In the present study we 
complement these results by exploiting the flexibility of the ETB formalism where strain effects may be entirely removed from the model Hamiltonian (without any structural simplifying assumption), allowing direct comparison of the real QD with an artificially strain-unaffected GaAs-embedded QD.
\item Analyze the influence of strain field and inter-dot hybridization for well separated dots. Previous studies
of inter-dot coupling \cite{vasanelli,pryor2} focused on a complementary range of dot separations, namely closely stacked
dots.
\end{enumerate}

We calculate the single-particle bound electron and hole states and wave functions adopting Boykin's 
 $sp^{3}s^{*}$ parametrization with interactions up to 2nd nearest neighbors and spin-orbit coupling. \cite{boykin} This parametrization gives very good fits of the important effective masses and gaps for bulk GaAs and InAs. 
 One potential problem is that it does not reproduce as well the $d$-bands contributions as parametrizations that explicitly
include $d$-orbitals, as for instance the one proposed by Jancu \emph{et al.}, \cite{bassani} where a $sp^{3}d^{5}s^{*}$ basis set and 1st nearest neighbor interactions were considered. Since the electron bound states in our QD come mainly from $s$ and $p$ atomic states, this does not 
constitute a relevant drawback. 
We do not consider piezoelectric effects in our model. However, as we will remark in Sec.~\ref{results_strained}, we expect 
only minor corrections to our results coming from such effects.
The influence of strain on the bond lengths is taken into account through a power-law scaling of the ETB parameters 
chosen here as such as to reproduce hydrostatic pressure effects in both bulk materials, while the influence of strain on the bond angles is taken into account by a generalized Slater-Koster formalism. \cite{slater}

This paper is organized as follows. In Sec.~\ref{methods} we present the formalism adopted for structural relaxation
of the system as well as for the electronic structure calculations, including the geometrical power-law scaling of
the ETB parameters.
Our results are given in Sec.~\ref{results}, and in Sec.~\ref{summary} we present a summary and conclusions.

\section{Formalism}
\label{methods}

\subsection{Structural analysis}
\label{methods_shape}

In order to calculate the atomic structure of an InAs quantum dot embedded 
in a GaAs matrix, i.e. the strain relaxation, two different methods have been used in the literature. One approach is an extrapolation of continuum nonlinear elasticity (CE) theory to the atomic scale, employing a discretization which is either based on the 
finite differences \cite{pryor,stier} or the finite-element (FE) method. \cite{liu99,moll98} The alternative is the valence-force field (VFF) approach, in particular the Keating model. \cite{keating,martin,williamson2}
The latter approach has several advantages : it accounts for internal displacements between the two sublattices of a zincblende crystal, that cannot be addressed within conventional continuum elasticity theory, and gives a displacement field which obeys the correct symmetry group $C_{2v}$. \cite{pryor1}
However, for large systems and slowly varying strain fields, the computational effort using the VFF approach is higher than using the FE method because, in the FE calculation, atomic resolution is usually not required in all regions of space.
In the present study, we started from the continuum elasticity theory as implemented in the FE method (using experimental elastic constants \cite{lb79}) to get a 
first approximation to the displacement fields. 
Then, by interpolation for the atomic positions that lie between the nodes of the FE's, we extracted all the positions of the atoms in the supercell that give rise to the calculated displacement field.
Finally we refined these 
displacements by further relaxing the atomic positions using a VFF model. 
Thus the positions of the atoms are eventually determined within the VFF model, 
thereby ensuring the correct $C_{2v}$ symmetry of the atomic displacements fields.

In the VFF model, the elastic energy energy of a zincblende lattice is expressed as a function of the atomic positions 
\{$\mathbf{R}_{\mathit{i}}$\} as

\begin{eqnarray}
E & = & \sum_{i} \sum_{j=1}^{4} \frac{3\ \alpha_{ij}}{16 \left( d^{0}_{ij} \right)^{2}}
\left[ \left( \mathbf{R}_{\mathit{j}} -\mathbf{R}_{\mathit{i}} \right)^{2} - \left( d^{0}_{ij} \right)^{2}  \right]^{2}  
\nonumber \\
& & + \sum_{i} \sum_{j,k>j} \frac{3\ \beta_{ijk}}{8\ d^{0}_{ij} d^{0}_{ik}} 
\left[ \left( \mathbf{R}_{\mathit{j}} - \mathbf{R}_{\mathit{i}}  \right) \cdot
\left( \mathbf{R}_{\mathit{k}} - \mathbf{R}_{\mathit{i}}  \right) \right. 
\nonumber \\
& & \left. - \cos \theta_{0}\ d^{0}_{ij} d^{0}_{ik} \right]^{2}.
\label{vff}
\end{eqnarray}
Here, $d^{0}_{ij}$ denotes the bulk equilibrium  bond length between nearest neighbor atoms $i$ and $j$ in the corresponding
binary compound, and $\theta_{0}=\cos^{-1}(-1/3)$ 
is the ideal bond angle. The first term is a sum over all atoms $i$ and its four nearest neighbors $j$, the second term is a sum over all atoms $i$ and 
its distinct pairs of nearest neighbors $j$ and $k$. The 
local-environment-dependent coefficients, $\alpha_{ij}$ and $\beta_{ijk}$ are the bond-stretching and bond-bending force 
constants respectively. We use \cite{pryor1} for GaAs $d_{ij}^{0}=2.448$ \AA, $\alpha_{ij}=41.49 \times 10^{3}$ 
dyne/cm, $\beta_{ijk}=8.94 \times 10^{3}$ dyne/cm; for InAs we use $d_{ij}^{0}=2.622$ \AA, $\alpha=35.18 \times 10^{3}$ 
dyne/cm, $\beta_{ijk}=5.49 \times 10^{3}$ dyne/cm. Across the heterointerfaces, where the species $j$ and $k$ are different 
(Ga and In), we use for $\beta$ the geometrical average of the corresponding values for pure GaAs and InAs.

The elastic energy is minimized with respect to the atomic positions \{$\mathbf{R}_{\mathit{i}}$ \}. 
In the minimization process, each atom is moved along the direction of the 
force on it, \mbox{$\mathbf{F}_{\mathit{i}}=-\nabla_{i} E$}, and the movement is iterated until this force is smaller than 
0.001 eV/\AA.

We compared the elastic constants derived from the VFF model to the experimental ones. The elastic constants $C_{11}$ and $C_{12}$ agree with the experimental values within 6 \%. Differences are noticeable mainly in $C_{44}$. The VFF model gives the $C_{44}$ about 10 \% too low for GaAs and about 20 \% too low for InAs.
In order to estimate the error due to inaccurate elastic constants, we calculated the local strain tensor by CE using both the elastic constants derived from the VFF model and the experimental ones. By comparing these results we verified that the absolute error in the diagonal components of strain tensor was always smaller than 0.005.

\subsection{Electronic calculations}
\label{methods_electronic}

The electronic structure is 
obtained within the ETB approach, adopting a $sp^{3}s^{*}$ parametrization with interaction up to 2nd nearest neighbors  
and spin-orbit coupling, \cite{boykin} which has been successfully used for III-V semiconductor heterostructures. 
\cite{menchero1,menchero2}
The wave functions are written as 
$\Psi=\sum_{i \nu \sigma} c_{i \nu \sigma} |i \nu>_{\sigma}$, where $|i \nu>_{\sigma}$ are orthogonal normalized $\mathbf{R}_{\mathrm{i}}$-centered orbitals of angular  
type $\nu=s,p_{x},p_{y},p_{z},s^{*}$ and spin $\sigma$, and 
$c_{i\nu \sigma}$ are complex expansion coefficients. 
The $s^{*}$ orbital was first introduced by Vogl 
\textit{et al.}\cite{vogl} to obtain a better description of the conduction bands.
In a $N$-atoms system, the $10 N \times 10 N$ ETB Hamiltonian matrix contains 33 independent matrix elements for bulk GaAs 
and 33 for bulk InAs. These matrix elements are the parameters of the model for the present calculation, and are taken from Ref.~\onlinecite{boykin}
In a strained InAs/GaAs mixed material, such as the QD system, a new parameter related to the valence band offset also needs to be included in the model. This parameter consists in a shift of all diagonal Hamiltonian matrix elements for bulk InAs (resulting in an analogous shift of the InAs bands), and it has been chosen such that the energy difference between the bulk InAs and the bulk GaAs valence band edge to coincide with the bulk valence band offset ($\Delta_{v}$).  

We performed an analysis of the QD gap dependence on the specific choice of $\Delta_{v}$, because there is a considerable spread in the experimental values  reported for $\Delta_{v}$ in the literature. \cite{vurgaftman} By varying $\Delta_{v}$ in the range 52 - 300 meV we obtained a QD gap variation smaller than 4 \%, indicating that in this range our results are not much affected by the specific choice of the offset. In what follows we take $\Delta_{v}$=52 meV from Ref.~\onlinecite{zunger1}, in order to better compare with results reported there .

The relaxed geometry of the QD system implies changes in bond lengths and in bond angles as compared to the ideal bulk materials. Both effects are incorporated in our electronic model.
Bond length deviations with respect to the bulk equilibrium distances $d_{ij}^{0}$ introduce corrections to the ETB Hamiltonian off-diagonal elements $V_{kl}$. 
Note that recently a different scheme has been proposed, \cite{klimeck1} where corrections to the diagonal matrix elements have also been included, in the framework of the $sp^{3}d^{5}s^{*}$ 1st nearest neighbors parametrization. \cite{bassani}    
We assume a power-law scaling \cite{harrison} for the off-diagonal elements :

\begin{equation}
V_{kl} \left( \left| \mathbf{R}_{\mathrm{i}}-\mathbf{R}_{\mathrm{j}}\right| \right) = V_{kl}(d_{ij}^{0}) \ \left( \frac{ d_{ij}^{0} }{ \left| \mathbf{R}_{\mathrm{i}}-\mathbf{R}_{\mathrm{j}} \right| } \right)^{n_{kl}},
\label{scaling}
\end{equation}
where $\left| \mathbf{R}_{\mathrm{i}}-\mathbf{R}_{\mathrm{j}} \right|$ is the actual bond length and $V_{kl}(d_{ij}^{0})$ is the bulk matrix element taken from Ref.~\onlinecite{boykin} ($k$ and $l$
label the different matrix elements). The exponents $n_{kl}$ are determined to reproduce variations of the relevant binary 
materials electronic properties under hydrostatic pressure, namely the volume deformation potentials $a_{v}^{\alpha}$ 

\begin{equation}
a_{v}^{\alpha} = V \left. \frac{\partial \epsilon_{gap}^{\alpha}}{\partial V} \right|_{V_{0}},
\end{equation}
for $\alpha$ corresponding to the 
direct as well as indirect ($\Gamma-L$ and $\Gamma-X$) band gaps.

\begin{table*}
\begin{ruledtabular}
\begin{tabular}{l|c c c c|c c c c} 
& \multicolumn{4}{c|}{GaAs} & \multicolumn{4}{c}{InAs} \\
& Exp. & PP & LAPW & ETB & Exp. & PP & LAPW & ETB \\ \hline
$a_{v}^{\Gamma-\Gamma}$ & $-8.5 \pm 0.5$\footnote[1]{Ref.~\onlinecite{pfeffer}} & -8.33 & -8.15 & -8.2 & -6.0\footnotemark[1] & -6.08 & -5.66 & -6.1  \\ 
$a_{v}^{\Gamma-L}$ & -- & -- & -3.70 & -3.4 & -- & -- & -2.89 & -2.9 \\
$a_{v}^{\Gamma-X}$ & -- & -- & 1.05 & 0.4 & -- & -- & 0.92 & 0.2 \\ 
$a_{v}^{\Gamma_{6c}}$ & -- & -7.17 & -8.46 & -6.7 & -- & -5.08 & -5.93 & -5.1 \\
$b$ & -2.0 & -1.90 & -- & -1.7 & -1.8 & -1.55 & -- & -2.0 \\
$d$ & -5.4 & -4.23 & -- & -3.5 & -3.6 & -3.10 & -- & -3.1 \\ 
\end{tabular}
\end{ruledtabular}
\caption{\label{voldefpot}Volume deformation potentials (in $eV$) for direct ($a_{v}^{\Gamma-\Gamma}$) and indirect band gaps 
($a_{v}^{\Gamma-L}$ and $a_{v}^{\Gamma-X}$), absolute volume deformation potential for the conduction band edge 
($a_{v}^{\Gamma_{6c}}$) and deformation potentials for uniaxial strains along [001] ($b$) and along [111] ($d$) (see text). 
Uniaxial strains were applied starting from the experimental \cite{lb79} lattice constants. For the uniaxial strain along [111], 
the internal atomic displacement is calculated using the VFF method.
Our ETB calculations are compared with experimental results \cite{lb79} (Exp.), with DFT-LDA calculations using pseudopotentials (PP) \cite{vandewalle} and with DFT-LDA calculations using the LAPW method (LAPW). \cite{wei}}
\end{table*}

In Table \ref{voldefpot} we give the values for $a_{v}^{\alpha}$ for 
GaAs and InAs taken from experiments, \cite{lb79} from a density-functional theory (DFT) calculation using the local-density approximation (LDA) and \emph{ab initio} pseudopotentials (PP), 
\cite{vandewalle} from a DFT-LDA calculation with the linearized augmented planewave (LAPW) method,  \cite{wei} and from our results (ETB).
In principle each $n_{kl}$ depends on the orbital character. However, we find that a single exponent $n_{kl}=3.40$ for all integrals and both materials gives a satisfactory agreement with LAPW for $a_{v}^{\Gamma - \Gamma}$ and $a_{v}^{\Gamma - L}$. For  $a_{v}^{\Gamma-X}$ the agreement is less satisfactory. We believe that this difference reflects the fact that the bottom of the conduction band at $X$ has a noticeable $d$ contribution. \cite{bassani} Therefore the inclusion of $d$ states in the parametrization and a correspondent different $n_{kl}$ value (cf. Eq.~\ref{scaling}) would be necessary.
However for InAs/GaAs QD's at atmospheric pressure, the confinement effect for electrons and holes inside the dot comes from the conduction and valence band offsets at the $\Gamma$ point,~\cite{williamson1} whereas the $X$ point does not play an important role. Therefore we do not expect this disagreement to be relevant in our calculations. 
It is interesting to note that this single exponent is very close to the value
$n=3.454$ reported for GaAs and AlAs within a different ETB parametrization. \cite{capaz}
In Table \ref{voldefpot} we also give the absolute volume deformation potential for the conduction band edge ($a_{v}^{\Gamma_{6c}}$), and the deformation potentials for uniaxial strains along [001] ($b$) and along [111] ($d$), obtained by \cite{vandewalle}

\begin{eqnarray}
\delta \epsilon_{001} & = &  3 \ b \ (\epsilon_{zz}^{001} - \epsilon_{xx}^{001} ) \nonumber \\
\delta \epsilon_{111} & = &  3 \sqrt{3} \ d \ \epsilon_{xy}^{111}
\end{eqnarray}
where $\delta \epsilon_{001}$  and $\delta \epsilon_{111}$ are the energies of the light hole band with respect to the heavy hole band,
 in the absence of spin-orbit coupling, for strains along [001]  and [111] respectively, and $\epsilon_{ij}^{001}$ and 
$\epsilon_{ij}^{111}$ are components of the correspondent strain tensors, defined as

\begin{eqnarray}
\epsilon_{ij}^{001} & = \ \epsilon &  \left(  
\begin{array}{ccc}
-\frac{C_{12}}{C_{11} + C_{12}} & 0 & 0 \\ 
0 & -\frac{C_{12}}{C_{11} + C_{12}} & 0 \\
0 & 0 & 1
\end{array}
                                      \right) \nonumber \\
\epsilon_{ij}^{111} & = \ \epsilon &  \left(
\begin{array}{ccc}
\frac{2 C_{44}}{C_{11} + 2 C_{12}} & 1 & 1 \\
1 & \frac{2 C_{44}}{C_{11} + 2 C_{12}} & 1 \\
1 & 1 & \frac{2 C_{44}}{C_{11} + 2 C_{12}}
\end{array}
                                      \right)
\end{eqnarray}
where $C_{11}$, $C_{12}$ and $C_{44}$ are the experimental \cite{lb79} elastic constants. 
For the uniaxial strain along [111], the internal atomic displacement is calculated using the VFF method.

\begin{figure}
\resizebox{85mm}{!}{\includegraphics{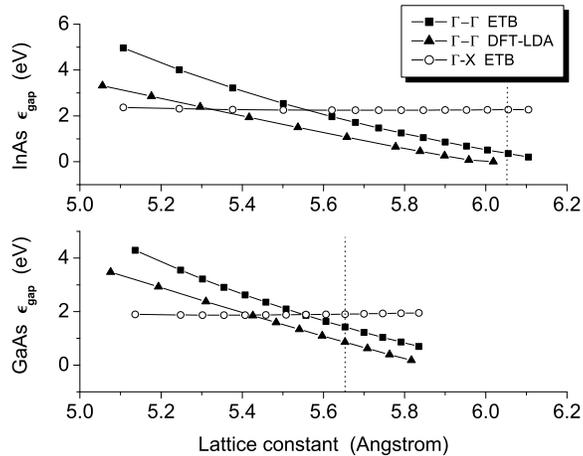}}
\caption{\label{bandgap_behav_comp}Comparison between ETB and DFT-LDA results for the InAs and GaAs $\Gamma - \Gamma$ band gaps obtained by varying the lattice constant. We also report the $\Gamma - X$ band gap calculated by ETB. The vertical dashed lines mark the bulk lattice constants (6.055~\AA\ for InAs and 5.653~\AA\ for GaAs).  }
\end{figure}

Fig.~\ref{bandgap_behav_comp} confirms the adequacy of the single-exponent scaling for the present study by comparing our ETB with DFT-LDA results for the InAs and GaAs $\Gamma-\Gamma$ band gaps obtained by varying the lattice constant. 
The DFT-LDA calculations were performed using scalar-relativistic \emph{ab initio} pseudopotentials of the Hamann type. \cite{fuchs} The electronic wave functions were expanded into a plane-wave basis set with a cut-off energy of 16 Ryd.
It is well known that the band gap is underestimated in LDA, but the overall behavior of the gap vs. hydrostatic lattice deformation should be reliable. Our ETB scheme yields the experimental optical band gap and the volume deformation dependence follows closely the trend obtained
with the LDA for a wide range of deformations.
On the same figure, we also reported the $\Gamma - X$ band gap calculated by ETB. We can observe that the band at the $X$ point has higher energy than at the $\Gamma$ point, for the whole range of lattice distortions typical in a QD (where the InAs (GaAs) is compressed (expanded) by at most 7~\%). This behavior would not change if the ETB-calculated $a_{v}^{\Gamma - X}$ reproduced better the LAPW results (see Table \ref{voldefpot}), since the curve representing the $\Gamma - X$ band gap has a slope much smaller than the $\Gamma - \Gamma$ band gap curve, and the crossing point between them would not change its position appreciably. These considerations confirm that a more accurate $a_{v}^{\Gamma - X}$ would not affect the results presented here.

Bond angle distortions are included in the ETB Hamiltonian as suggested in the Slater-Koster formalism, \cite{slater} generalized to 
include three-center integrals for the 18 independent 2nd nearest-neighbor matrix elements. \cite{comment}
Note that, different
from previous studies \cite{tserbak}, we do not assume that the three-center integrals are independent of directional 
changes induced by the strain.

The relevant eigenstates of the resulting Hamiltonian matrix including all the strain effects,$\hat{H}$, are obtained variationally. We build the quotient

\begin{equation}
\Re \left[ \varphi \right] = \frac{\left\langle \varphi \left| \left( \hat{H}-\epsilon_{r} \hat{I} \right)^{2} \right| \varphi \right\rangle}
{\left\langle \varphi | \varphi \right\rangle},
\end{equation}
\noindent where $\epsilon_{r}$ is a reference energy. By minimizing $\Re$ with respect to the trial function $\varphi$ by a steepest descent 
algorithm, we get the eigenvector (and the related eigenvalue) whose energy is nearest to $\epsilon_{r}$. Therefore by 
varying $\epsilon_{r}$ we may in principle determine all the electron and hole bound state energies and wave functions.\cite{capaz,menchero1,menchero2,wang3}

Within our ETB formalism, strain effects can be formally removed from the electronic calculation by imposing $n_{kl}=0$ in 
Eq. (\ref{scaling}) (removal of the strain and relaxation effects from bond lengths) and setting the direction cosines between atomic 
orbitals equal to the corresponding bulk values (removal of the strain and relaxation effects from bond angles). Therefore, by contrasting the bound state energies of an artificially strain-unaffected 
QD with the corresponding results for the strained QD, we are able to quantify the total strain impact  on the electronic properties.

\section{Results}
\label{results}

\subsection{Relaxed QD geometry}
\label{strained_QD_geometry}

\begin{figure}
\resizebox{85mm}{!}{\includegraphics{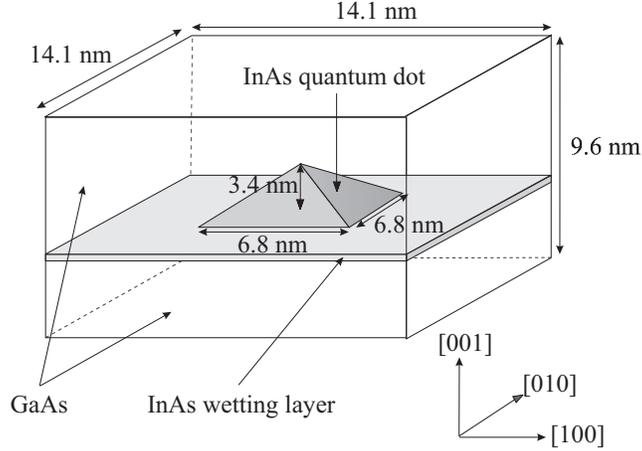}}
\caption{\label{pyramid}Schematic view of the pyramidal InAs QD buried in the GaAs matrix. The wetting layer is one monolayer- 
thick. The pyramid base length is about 6.8 nm, the height is about 3.4 nm. The supercell contains 85 000 atoms.}
\end{figure}

Fig.~\ref{pyramid} is a schematic view of our pyramidal InAs QD buried in a GaAs matrix. The wetting layer is modeled by a monolayer-thick InAs layer at the base of the pyramid. 
The pyramid base length is $12a$, the height is $6a$, where $a$=5.653 \AA \ is the lattice constant of bulk zincblende GaAs. 
We place the InAs pyramid and wetting layer in a large GaAs box, to which periodic boundary conditions are applied. This supercell is used in our structural and electronic calculations. Three different supercells were considered, containing the same QD but differing by the size of the GaAs matrix, namely GaAs matrices with dimensions $19a \times 19a \times 14.067a$ (40~432 atoms), $25a \times 25a \times 17.067a$
(85~000 atoms - shown in Fig.~\ref{pyramid}), and $37a \times 37a \times 35.067a$ (383~320 atoms).
The $z$-dimension is not an integer multiple of $a$ due to the InAs wetting layer. 
Unless specified otherwise, results presented below refer to the 85 000 atoms supercell.

\begin{figure*}
\resizebox{85mm}{!}{\includegraphics{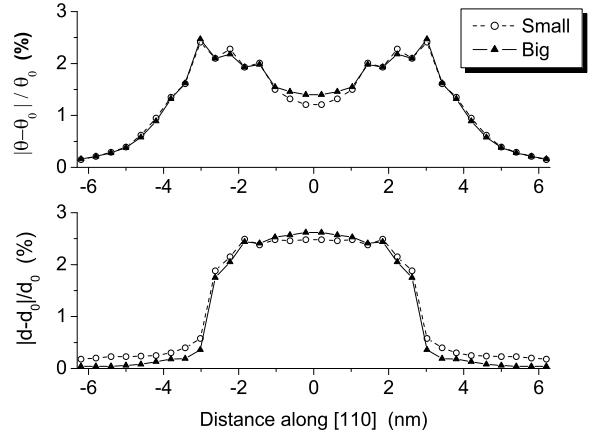}}
\caption{\label{strain_110} Average relative distortion of the bond angle $\theta$ from the ideal zincblende bond angle $\theta_{0}$, and average
relative distortion of the bond length $d$ from the ideal GaAs and InAs bond lengths $d_{0}$. These quantities are calculated along the [110] direction at $z=0.4\ h$. For comparison we present results obtained from the 40 432-atoms supercell (Small) and the 383 320-atoms supercell (Big).   }
\end{figure*}

In Fig.~\ref{strain_110} we present the relative distortion of the bond angle $\theta$ from the ideal zincblende bond angle $\theta_{0}$ (obtained by averaging over the 6 different bond angles around each cation), and the 
relative distortion of the bond length $d$ from the ideal InAs (inside the dot) and GaAs (outside the dot)  bond lengths $d_{0}$ (averaging over the 4 bond lengths).
These quantities are calculated along the [110] direction at $z=0.4\ h$, where $h$ is the QD high.  We compare results obtained from the 40 432-atoms supercell (Small) and the 383 320-atoms supercell (Big).
The crystallographic directions are defined by starting from a zincblende unit cell containing a cation at the origin and an anion at $a(\frac{1}{4},\frac{1}{4},\frac{1}{4})$. 
The figure shows that the thicker GaAs region allows for a better relaxation of the bond lengths in the largest part of the supercell, 
while it is not as efficient in angular relaxations.

\subsection{Bound states of the relaxed QD}
\label{results_strained}

\begin{figure*}
\resizebox{85mm}{!}{\includegraphics{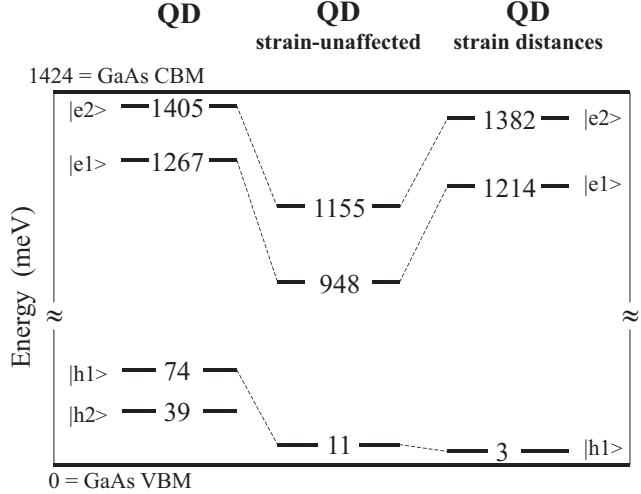}}
\caption{\label{energies_shift} QD bound state energies (in meV) calculated by our ETB approach. The energy zero is the bulk GaAs valence band maximum. Different degrees of strain are taken into account. In the first column (QD) the bound state energies of the {\em physical} QD are reported, where the strain effects have been included in the ETB hamiltonian. The second column (QD strain-unaffected) gives these energies for an artificially strain-unaffected QD, discussed in Sec.~\ref{results_unstrained}. The third column (QD strain distances) gives results when strain is retained only in the bond length description, while bond angles are assumed to equal the bulk ones (discussed in Sec.~\ref{results_unstrained}).   }
\end{figure*}

We have calculated electron and hole bound states, and refer to them as $|e1>$ and $|h1>$ for the respective ground states, $|e2>$ and $|h2>$ for the next excited states, and so on. For the QD supercell containing 85 000 atoms the energies are calculated as 1405 meV ($|e2>$ state), 1267 meV 
($|e1>$ state), 74 meV ($|h1>$ state) and 39 meV ($|h2>$ state) using the top of the bulk GaAs valence band as energy zero.
These energies are shown in Fig.~\ref{energies_shift}, on the left side (column labeled QD).
 From our numerical approach, we cannot exclude the possible existence of other hole states with 
smaller energies and very close ($\Delta \epsilon < 10-15$ meV) to the $|h2>$ state.

\begin{figure*}
\begin{center}
\resizebox{175mm}{!}{\includegraphics[33mm,90mm][210mm,260mm]{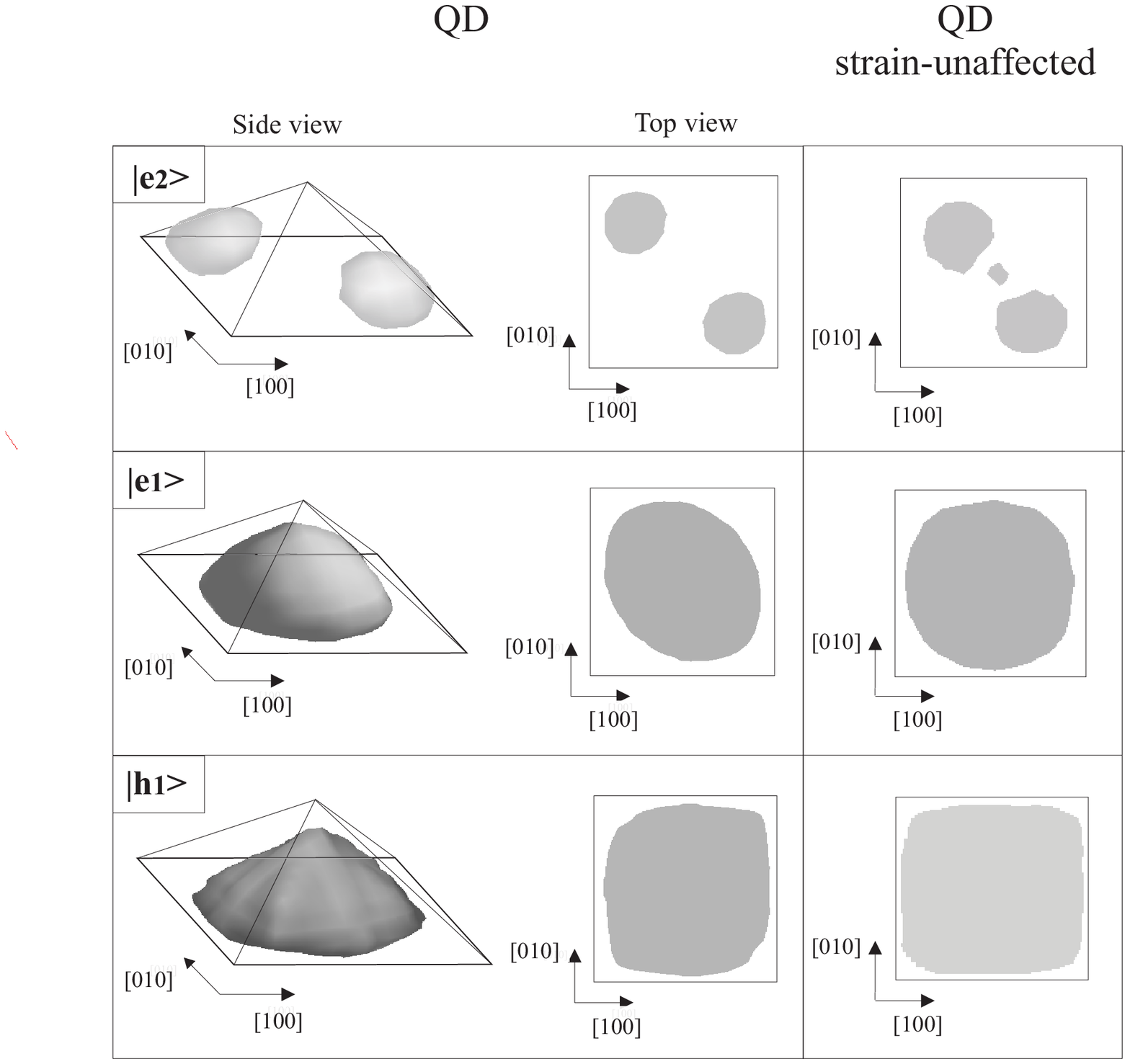}}
\caption{\label{wavefunctions}Isosurface plots of the charge densities $e |\varphi (\mathbf{r})|^{2}$ relative to the electron states $|e2>$ and $|e1>$, and to the hole state $|h1>$. Each 
surface correspond to 0.5 of the maximum charge density value. The left side (QD) shows the results obtained from the {\em physical} QD, while the right side (QD strain-unaffected) shows the results obtained from the strain-unaffected dot (discussed in Sec.~\ref{results_unstrained}).}
\end{center}
\end{figure*}

We show in Fig.~\ref{wavefunctions}, on the left side (QD), the isosurface plots of the charge densities $e |\varphi (\mathbf{r})|^{2}$ corresponding to the electron states $|e2>$ and $|e1>$, and to the hole state $|h1>$. The isosurfaces are selected as 0.5 of the maximum charge density value. 
The figure
shows that the charge is almost entirely confined inside the dot.
The lowest electron state ($|e1>$) is almost $s$-like (slightly elongated along [$\overline{1}$10]), the next electron state ($|e2>$) is $p$-like 
aligned along [$\overline{1}$10], and the hole state ($|h1>$) has an elongation perpendicular to the $|e1>$ state, in agreement with the work of Stier~\emph{et al.}\cite{stier} and Wang~\emph{et al.}\cite{zunger1} 
In Table \ref{gaps} we show a comparison of the energy differences between the QD bound states calculated within the empirical pseudopotential approach (PP) (whose results are extracted from Fig. 2 of Ref.~\onlinecite{zunger1}) and the present  approach (ETB strained). 
We can see that the agreement is good.

\begin{table}
\begin{ruledtabular}
\begin{tabular}{ccc}
& PP & ETB \\ \hline
$\epsilon_{|e2>} - \epsilon_{|e1>}$ & 130 & 138 \\
$\epsilon_{|e1>} - \epsilon_{|h1>}$ & 1150 & 1193 \\
$\epsilon_{|h1>} - \epsilon_{|h2>}$ & 25 & 35 \\
\end{tabular}
\end{ruledtabular}
\caption{\label{gaps}Comparison of the QD bound state energy differences (in meV) obtained from empirical pseudopotentials \cite{zunger1} (PP) (see text) and the present ETB results (ETB). }
\end{table}

In larger pyramidal QD's, an additional $p$-like electron state oriented perpendicularly to $|e2>$  is usually present.~\cite{stier,zunger1} When calculated within macroscopic models, these two $p$-like states are degenerate, if the piezoelectric effects are neglected.~\cite{stier}
In our calculations we did not find this additional $p$-like state, and we observed that the $|e2>$ state lies about 20 meV below the GaAs conduction band edge. On the other hand, when we considered the artificial strain-unaffected QD (see next section), all the electron states became deeper, and the additional $p$-state did appear about 30 meV above $|e2>$. 
This degeneracy lifting appears in our atomistic model 
as a consequence of the breaking of the pyramidal $C_{4v}$ symmetry into the lower zincblende $C_{2v}$ symmetry.

\begin{figure}
\resizebox{85mm}{!}{\includegraphics{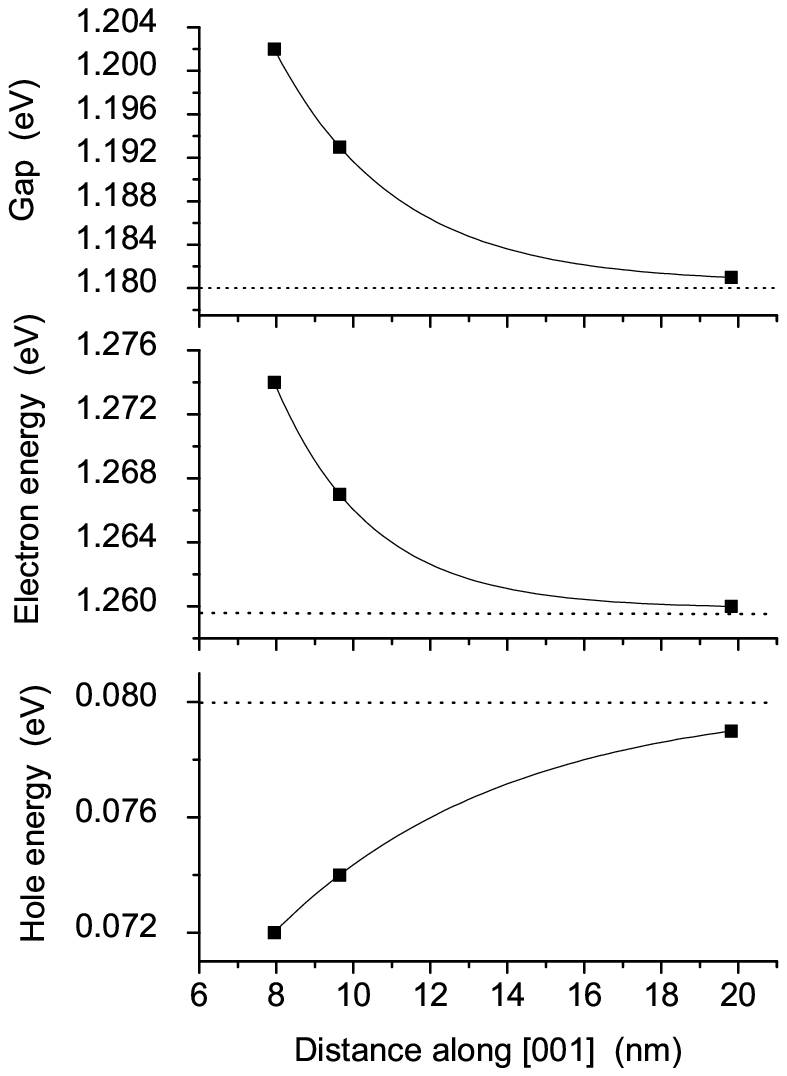}}
\caption{\label{energies_strain}Single-particle electron and hole energies ($\epsilon_{|e1>}$ and $\epsilon_{|h1>}$ respectively) and QD gap ($\epsilon_{|e1>} - \epsilon_{|h1>}$) as functions of the distance between dots along [001]. The solid curves are phenomenological exponential fits, while the horizontal dashed lines are their asymptotes.}
\end{figure}

We now analyze how the first electron and the first hole states are affected by the supercell size. Due to the periodic boundary conditions, different cell sizes correspond to periodic three-dimensional QD arrays with different inter-dot separations. In each case, before performing the electronic calculation,  the atomic positions are relaxed as described in Sec.~\ref{methods_shape}. Supercell-size effects are due to electronic and elastic dot-dot interactions. 
Both contribute to the results shown in Fig.~\ref{energies_strain}, where the energies for the states $|e1>$ and $|h1>$ and the QD gap are shown for the three different supercells. 
The horizontal axis represents the supercell dimension along [001], i.e. the base-to-base inter-dot distance along the [001] direction. 
Although we have chosen to report our results here and in the following section as a function of the inter-dot 
distance along this direction, dot-dot interaction effects in all directions in the three-dimensional QD array are included.
We note that the dot-dot coupling in the 85 000-atoms supercell makes the gap wider by about 12 meV with respect to an isolated dot.
Strictly speaking, when we bring QD's together to form a periodic array, the bound states of the isolated dot spread into minibands whose width increases with the dot-dot interaction. 

Now a brief remark on the possible influence of the piezoelectric effects on the results shown in Fig.~\ref{energies_strain} (and in the subsequent Fig.~\ref{energies_nostrain}, in the next section). Grundmann \emph{et al.}~\cite{grundmann} have shown that  the piezoelectric potential inside a pyramidal InAs/GaAs QD's has a quadrupole-like character in the [001] plane. Moreover Fig.~\ref{wavefunctions} shows that our $|e1>$ and $|h1>$ states are almost symmetric under rotations around a [001] axis passing through the tip of the pyramid. It follows that in the framework of nondegenerate perturbation theory (where the piezoelectric potential is taken as perturbation), the first-order corrections to the energies $\epsilon_{|e1>}$ and $\epsilon_{|h1>}$ almost vanish. We therefore expect that the inclusion of the piezoelectric effects would not strongly affect the results shown in Figs.~\ref{energies_strain} and \ref{energies_nostrain}.

\subsection{Bound states of an artificially strain-unaffected QD}
\label{results_unstrained}

The influence of strain on the electronic properties was studied here by comparing the {\em physical} QD bound states with  
the corresponding artificially strain-unaffected QD states, as explained in Section \ref{methods_electronic}.
The second column in Fig.~\ref{energies_shift} (QD strain-unaffected) gives the bound state energies for the artificially strain-unaffected QD, that should be compared with results from the {\em physical} dot (QD). Note that the energy $\epsilon_{|h2>}$ for the strain-unaffected dot is not given, because the state $|h2>$ is unbound.  
We observe that strain increases the QD gap ($\epsilon_{|e1>} - \epsilon_{|h1>}$)
 by about 25 \%, raising it from the strain-unaffected value 937 meV to the value 1193 meV. 
This behavior comes mainly from the InAs main gap increase when the structure is compressed by the surrounding GaAs matrix, \cite{pryor} as can be seen in Fig.~\ref{bandgap_behav_comp} in the case of bulk InAs.
The change in the QD gap due to strain reported here agrees qualitatively with effective mass calculations 
in elliptic dots. \cite{fu}
We note that the band-gap variation with the strain (wider or narrower gap) is dominated by the $|e1>$ 
state position (shallower or deeper confinement).
 We observe that the strain has opposite effects on electron and hole states: electron states become shallower, approaching the conduction band edge, while hole states become deeper, moving far from the valence band edge.

The last column of Fig.~\ref{energies_shift} (QD strain distances) gives results obtained by retaining in the Hamiltonian
only the bond length deformations, assuming bulk bond angles. This shows that the electronic properties of the QD are mainly affected by deviations of the bond lengths from the respective bulk ones.
We note that the QD gap for the ``partially strained'' system (1211 meV) is larger than for the {\em physical} QD. In fact the hole level actually drops with the bond length compression, but it rises with the bond angle distortions resulting from the QD geometry. The angular contribution dominates, leading to the smaller gap  in the {\em physical} QD case. We also observe that the angular strain contribution is more important for $|h1>$ (71 meV) than for $|e1>$ (53 meV), since the former has wave function atomic components predominantly $p$-type, while the latter has wave function atomic components almost purely $s$-like (thus spherically symmetric).

The right hand side of Fig.~\ref{wavefunctions} (QD strain-unaffected) shows the isosurface plots of the charge densities of the strain-unaffected QD. By comparing with the results of the {\em physical} dot, we observe that the spatial orientation of $|e2>$  does not depend on the mesoscopic $C_{2v}$ symmetry (resulting from the strain field), but depends on the alternating interface structures of the four \{101\} facets (resulting from the microscopic zincblende structure). On the other hand, the spatial elongations of $|e1>$ and $|h1>$ depend only on the mesoscopic $C_{2v}$ symmetry of the strain field, and not on the alternating interface structures of the four \{101\} facets.

\begin{figure}
\resizebox{85mm}{!}{\includegraphics{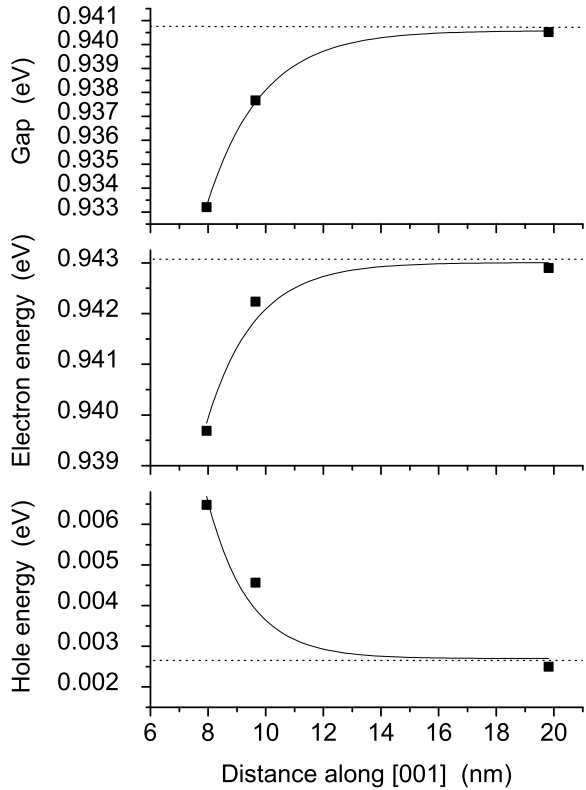}}
\caption{\label{energies_nostrain}Single-particle electron and hole energies ($\epsilon_{|e1>}$ and $\epsilon_{|h1>}$ respectively) and QD gap ($\epsilon_{|e1>} - \epsilon_{|h1>}$) as functions of the distance between dots along [001], in the case of artificially strain-unaffected QD's. The solid curves are phenomenological exponential fits, while the horizontal dashed lines are their asymptotes. Observe the different vertical scale from Fig.~\ref{energies_strain}.}
\end{figure}

In Sec.~\ref{results_strained} we discussed the supercell-size effects on the electronic and elastic dot-dot 
interactions. The results shown in Fig.~\ref{energies_strain} reflect the effects of both interactions. By repeating now the calculation for the $|e1>$ and $|h1>$ states in the three strain-unaffected supercells, we were able to isolate the electronic interaction due to the wave functions overlap. The results are shown in Fig.~\ref{energies_nostrain}, where the solid curves are phenomenological exponential fits and the horizontal dashed lines their asymptotes. 
All the fits are of the form $\epsilon = \epsilon_{0} + A \cdot \exp (-d/ \lambda)$, where $d$ is chosen as the base-to-base inter-dot distance along [001], $\epsilon_{0}$ represents the energy of the isolated dot (given by the horizontal line), $\lambda$ is the characteristic length of the interaction  along [001], and $A$ is a prefactor related to the interaction. 
For the overlap contribution (Fig.~\ref{energies_nostrain}), an exponential dependence is to be expected since this is the 
typical behavior of the localized wave functions away from the dot. For the strain field contribution, a power law dependence would be more realistic. \cite{bimberg1} We use exponential fits to allow a semi-quantitative comparison of an overlap-only case (Fig.~\ref{energies_nostrain}) with a situation where both effects are present (Fig.~\ref{energies_strain}). 
By considering the gap behavior, we obtain $A_{1} \approx 300$ meV and $\lambda_{1} \approx 3$ nm for the case considered in Fig.~\ref{energies_strain}, and $A_{2} \approx $ -500 meV and $\lambda_{2} \approx 2$ nm for the case in Fig.~\ref{energies_nostrain}.
Fig.~\ref{energies_nostrain} clearly shows the miniband broadening effect as the inter-dot distance is reduced, since the states here represented are, strictly speaking, the $|e1>$ miniband minimum and the $|h1>$ miniband maximum.
From the analysis of these results we come to the following conclusions regarding dots separated by distances 
which are at least a factor of 2 larger than the corresponding dot dimension: 

\begin{enumerate}
\item The range of the strain field interaction between dots is larger than that of the wave function overlap region.
\item In all general trends shown here, strain effects override direct wave function overlap effects, leading to 
the opposite behavior of the calculated energy variations versus distance. 
The net strain contribution to $\epsilon_{|e1>}$ and $\epsilon_{|h1>}$ would correspond to the subtraction of the data given in Fig.~\ref{energies_nostrain} from the corresponding frames in Fig.~\ref{energies_strain}.
Fig.~\ref{energies_strain} shows that the electron level downshifts when the inter-dot distance increases, while the hole level rises.
This behavior comes from the better relaxation of the bond lengths with the thicker GaAs region, shown in Fig.~\ref{strain_110}. 
This gives rise to a smaller bond-length component of the strain in the ETB Hamiltonian, and then, according to Fig.~\ref{energies_shift}, to a smaller electron energy and a larger hole energy.
\item The miniband width for both the $|e1>$ and $|h1>$ minibands is less than a few meV, consistent with 
the largest island spacings considered by Pryor, \cite{pryor2} where a rather different dot geometry and range of distances have been investigated.
\end{enumerate}

\section{Conclusions}
\label{summary}

We have generalized a previous ETB 2nd nearest neighbors parametrization by Boykin \cite{boykin} to include the lattice distortion into the Hamiltonian. We introduced a scaling law of 
the hopping Hamiltonian matrix elements with exponent $n=3.40$. We were able to reproduce the volume deformation
potentials corresponding to the direct ($\Gamma-\Gamma$) and the indirect ($\Gamma-L$) band gaps for both InAs and GaAs .
We have used this approach to calculate the electronic structure of a square based pyramidal 
quantum dot.
The comparison with previous empirical pseudopotential calculations shows that the ETB model provides accurate results 
for bound state energies and corresponding wave functions. 

The influence of strain on the bound state energies is analyzed. For single dots we found the strain increases the main gap by about 25~\%. Strain causes the electron states to become shallower and the hole states to become deeper.
The spatial orientation of the first $p$-state ($|e2>$)  depends on the alternating interface structures of the four \{101\} facets, while the spatial elongations of the ground electron ($|e1>$) and hole state ($|h1>$) depend on the mesoscopic $C_{2v}$ symmetry of the strain field. 

We have quantitatively discussed the influence of the dot-dot interaction on the bound states due to both strain field and wave function overlap by decoupling these two effects. 
For well separated dots, we have shown that the strain field dominates the level shifts, leading to opposite trends as for pure wave function overlap, although at distances less than twice the dot diameter the latter becomes noticeable.
The QD gap between the electron and hole states decreases as the GaAs region between dots gets thicker because this allows the bond lengths to further relax.

\begin{acknowledgments}
This work was partially supported by the Brazilian agencies CNPq, FAPERJ, Instituto do Mil\^{e}nio de Nanoci\^{e}ncias-MCT 
and by the Deutsche Forschungsgemeinschaft, SFB 296. 
We acknowledge E. Penev for contributing DFT results and A. S. Martins for useful discussions.
\end{acknowledgments}

\bibliography{biblio}

\begin{thebibliography}{51}
\bibitem[{com()}]{comment}
\bibinfo{note}{The three center integrals involve hopping matrix elements
  between orbitals centered at 2nd neighbor sites (A and B) whose values depend
  on the position of the common nearest neighbor site (C) in the zincblende
  structure. For distorted configurations, the relative displacement between A
  and B has been taken into account in the direction cosines \cite{slater},
  while the displacement of C with respect to A and B has been taken into
  account by interpolating linearly from the perfect structure matrix elements
  \cite{boykin} according to the deviation in the position of C with respect to
  the AB line.}
\end{thebibliography}

\end{document}